\documentclass[prb,superscriptaddress,showpacs,twocolumn]{revtex4-1}
\usepackage{graphicx}
\usepackage{amsmath,amssymb}
\usepackage{bm}
\usepackage[latin1]{inputenc}

\begin{document}

\title{Single-parameter pumping in graphene}

\author{Pablo San-Jose}
\affiliation{Instituto de Estructura de la Materia (IEM-CSIC), Serrano 123, 28006 Madrid, Spain}

\author{Elsa Prada}
\author{Sigmund Kohler}
\affiliation{Instituto de Ciencia de Materiales de Madrid (ICMM-CSIC), Cantoblanco, 28049 Madrid, Spain}

\author{Henning Schomerus}
\affiliation{Department of Physics, Lancaster University, Lancaster, LA1 4YB, United Kingdom}

\date{\today}

\begin{abstract}
We propose a quantum pump mechanism based on the particular
properties of graphene, namely chirality and bipolarity.  The
underlying physics is the excitation of evanescent modes entering a
potential barrier from one lead, while those from the other lead do
not reach the driving region.  This induces a large nonequilibrium
current with electrons stemming from a broad range of energies, in
contrast to the narrow resonances that govern the corresponding effect
in semiconductor heterostructures.  Moreover, the pump mechanism in
graphene turns out to be robust, with a simple parameter dependence,
which is beneficial for applications.  Numerical results from a
Floquet scattering formalism are complemented with analytical
solutions for small to moderate driving.
\end{abstract}


\maketitle

\section{Introduction}

Ratchets and pumps are devices in which spatio-temporal symmetry
breaking turns an ac force without net bias into directed
motion.\cite{Reimann2002a,Hanggi2009a}  If the time-dependence enters
via only one parameter, the pump current vanishes in the adiabatic
limit.\cite{Brouwer1998a}  Therefore, single-parameter pumping requires
non-equilibrium conditions enforced by driving beyond the adiabatic
limit.  This distinguishes single-parameter pumps from devices that
operate with oriented work cycles, like turnstiles \cite{Switkes1999a}
or sluices.\cite{Vartiainen2007a}
Quantum pumps can be implemented with quantum dots in a
two-dimensional electron gas (2DEG) driven by microwaves,
\cite{Oosterkamp1998a} surface-acoustic waves, \cite{Blumenthal2007a}
ac gate voltages \cite{DiCarlo2003a, Kaestner2008a, Fujiwara2008a,
Kaestner2009a} or non-equilibrium noise.\cite{Khrapai2006a}
Here we propose a single-parameter pump based on the particular
properties of graphene and show that these display a broad-band
response, in contrast to 2DEG-based ratchets or pumps for which isolated
resonances govern the effect.\cite{Oosterkamp1998a, Strass2005b,
Khrapai2006a}

\begin{figure}[b]
   \centering
   \includegraphics[width=8cm]{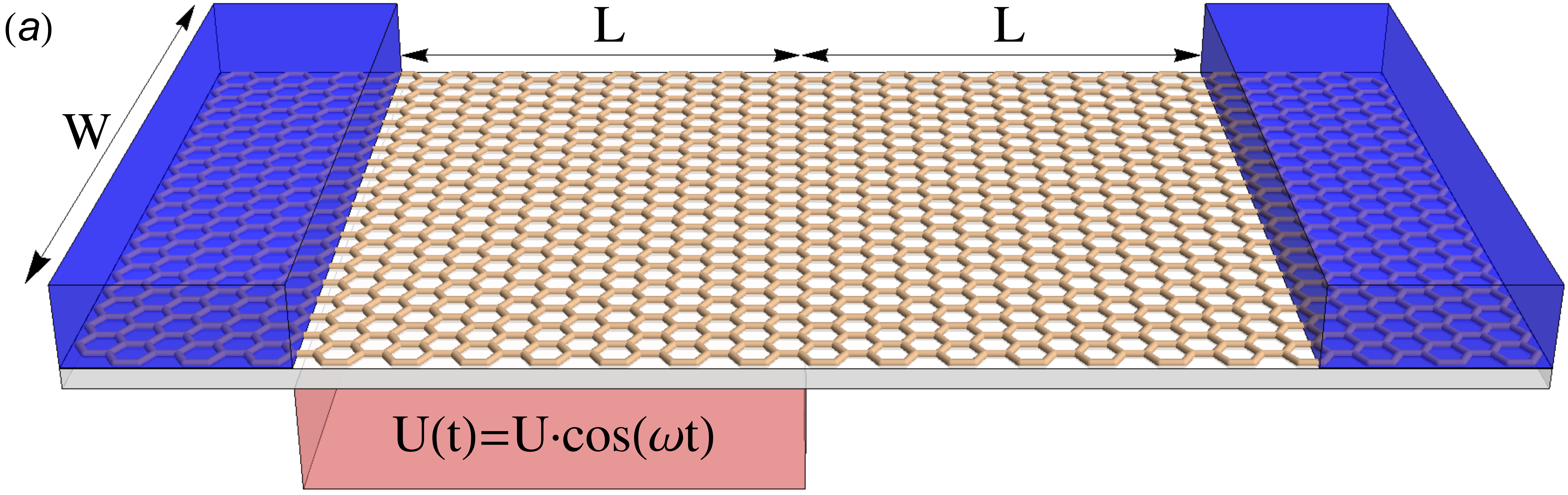}
   \\[2ex]
   \includegraphics[width=8cm]{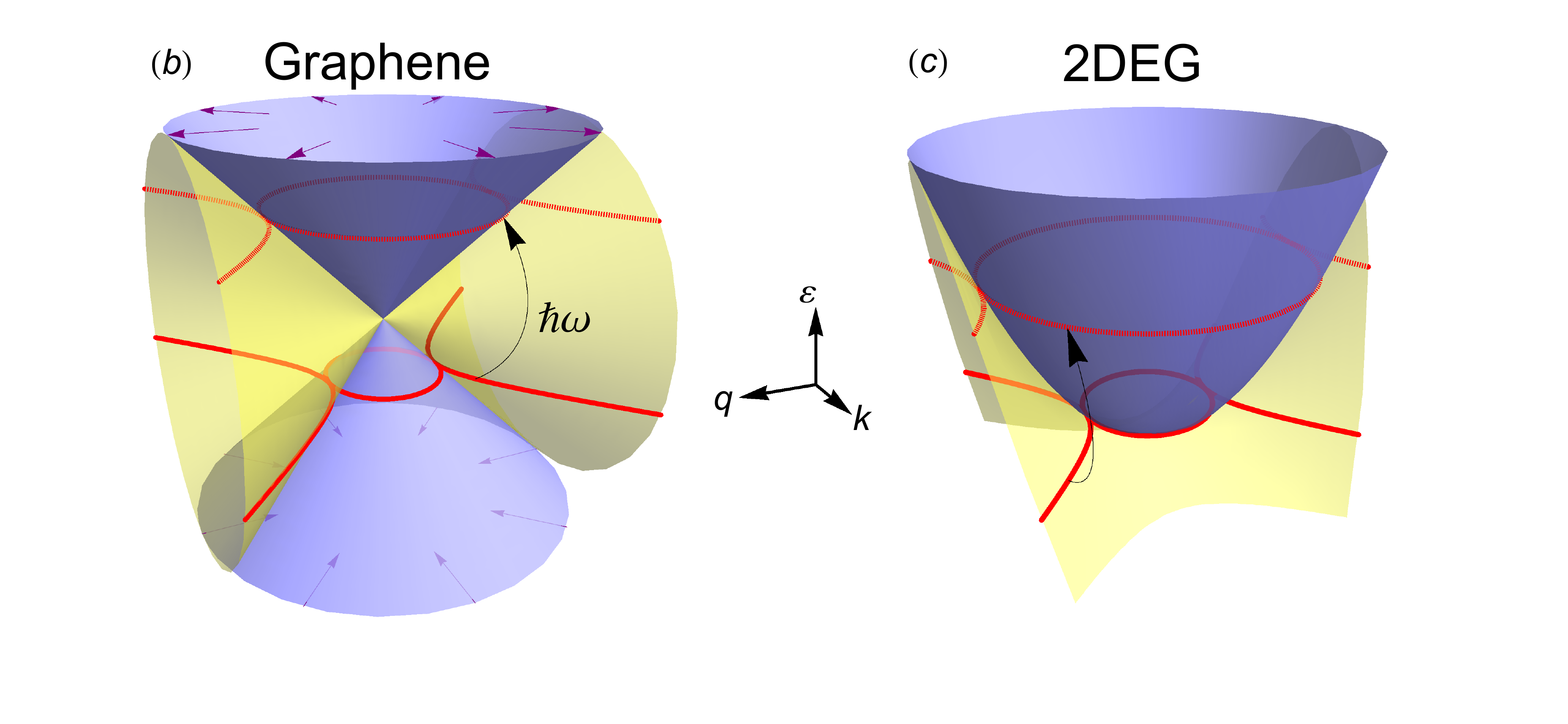}
   \caption{
Sketch of a single-parameter graphene pump connected to two leads.
The left half of the graphene ribbon is exposed to an ac gate voltage.
Dispersion of (b) graphene and (c) a 2DEG, with longitudinal and
transverse momenta $k$ and $q$, including the branches of evanescent modes.
Photon absorption (and in the case of graphene
also emission) may excite an evanescent mode to propagating.}
\label{Fig:system}
\end{figure}

Owing to the chiral and gapless nature of charge carriers in graphene, an
electron hitting a potential step in this material may propagate forwards as a
hole with opposite momentum.\cite{Klein1929a,Katsnelson2006a}
Due to this Klein tunneling it is difficult to confine electrons by
electrostatic potentials. This phenomenon even extends to evanescent modes,
i.e., modes that decay exponentially as a function of the barrier
penetration. In graphene, electrons populating such modes can tunnel a
large distance.\cite{Tworzydlo:PRL06} This impediment to
electrostatic confinement constitutes a drawback for many switching
and sensing applications.

An inspection of present quantum pump designs suggests that
graphene pumps would be negatively affected by this issue as well.
For example, pumps have been realized with electrostatically
defined double quantum dots in a 2DEG.\cite{Oosterkamp1998a,
Khrapai2006a}  In these experiments, a driving field induces dipole
transitions of an electron from a metastable state below the Fermi
energy in the, say, left dot to a metastable state in the right dot.
Subsequently the electron will leave to the right lead, and an
electron from the left lead will fill the empty state in the left dot.
The emerging pump current therefore requires resonant conditions
with a pair of energetically well-defined states, i.e., well isolated
states with long life times, which do not exist in graphene.

Nevertheless, the alternative, graphene-specific mechanism identified
here allows realizing highly efficient pumping, which may
indeed outperform conventional devices.  The large pump current
emerges in the bipolar regime around the Dirac point. This is due to a
scattering process where a whole continuum of evanescent modes is
promoted into unidirectionally propagating states, which couple well
to the leads because of chirality. Since this mechanism does not rely
on intricate resonance conditions, but invokes graphene's intrinsic
features of bipolarity and chirality, the pump current is robust
and displays a simple parameter dependence curve, as is desirable for
electronic sensor applications.

This paper is organized as follows: In the next section we describe
our proposal for a single-parameter pump and give typical values for
length and energies scales. In Sec.~\ref{Floquet} we introduce the
Floquet scattering theory that applies to non-equilibrium pumping in
graphene. We first present the general formalism and then consider the
weak driving regime, where we derive analytical expressions for the
one-photon transmission probability. In Sec.~\ref{current} we present
numerical results for the pumped current comparing the performances of
graphene and a 2DEG pump. Moreover, we compare the numerics with the
semiclassical approximation derived in the preceding section and find
good agreement. The specific pumping mechanism by excitation of
evanescent modes is explained in Sec.~\ref{direction}. Finally, we
conclude in Sec.~\ref{conclusions}.


\section{Description of the system}

As a single-parameter pump setup, sketched in Fig.~\ref{Fig:system}a,
we consider a graphene ribbon of length $2L$ and width $W$ attached to two metallic electron
reservoirs, with its left half driven by a time-dependent gate. The
driving by variation of only one parameter excludes the
emergence of an adiabatic pump current.\cite{Brouwer1998a,Prada:PRB09}
However, considering that spatial symmetry is broken by the placement
of the gate, a finite dc current arises in non-equilibrium conditions,
which can be achieved by non-adiabatic driving.  In order to assess
the importance of graphene's chirality and bipolarity in the
non-adiabatic context, we contrast our results with those found for
the corresponding setup of a 2DEG in a semiconductor heterostructure.

We assume that the sample sizes are smaller than the mean free path and that $W\gg L$, so that the effects coming from the boundaries, electron-electron interactions and disorder play a minor role. Besides, we restrict ourselves to the low energy regime such that the Dirac approximation remains
valid. Then, our quasi one-dimensional model contains three natural energy scales,
namely the driving frequency $\omega$, the driving amplitude $U$, and
the energy associated with the length $L$ of the device.  Additionally, the leads' Fermi momentum $\hbar k_F^{(\infty)}$ becomes a relevant scale in a 2DEG, since the contact resistance depends on its value. For graphene
with Fermi velocity $v_F \approx 10^6\mathrm{m/s}$ and $L=5\mu
\mathrm{m}$, the latter is $E_L^G=\hbar v_F/L \approx
0.13\mathrm{meV}$.  For a 2DEG with the same geometry and effective
mass $m^* = 0.067m_\mathrm{e}$ (GaAs/AlGaAs), it is roughly four
orders of magnitude smaller, $E_L^N=\hbar^2/2m^*L^2 \approx
0.02\mathrm{\mu eV}$.  Driving beyond the adiabatic limit requires
frequencies $\omega \gtrsim E_L/\hbar$.  Electrons can then absorb or
emit photons, and after a short transient period, these excitations will establish a non-equilibrium population of the electronic states. 


\section{Floquet scattering theory in graphene}\label{Floquet}


A quantitative description of nonadiabatic charge transport is
provided by Floquet scattering theory. Subsequently, we show that this
theory provides the probability $T^{(n)}_{LR}$ for an electron to be
scattered from the left to the right lead under the absorption or
emission of $n$ photons, and with it we can calculate the dc
current.\cite{Kohler2005a}  In order to focus on the graphene-specific
features, we will restrict ourselves to the zero temperature limit so
that all electronic states below the Fermi energy $E_F$ are initially
occupied.


\subsection{General formalism}


When scattered at a periodically time-dependent potential, an
electron with initial energy $\epsilon$ may absorb or emit $|n|$
quanta of the driving field ($n<0$ corresponds to emission), such that
its final energy is $\epsilon+n\hbar\omega$.  This is embodied in the
decomposition of the transmission probability from the left to the
right reservoir
\begin{equation}
T_{LR}(\epsilon) = \sum_{n=-\infty}^\infty T_{LR}^{(n)}(\epsilon) .
\end{equation}
The corresponding time-averaged dc current is given by the generalized Landauer
formula\cite{Wagner1999a,Kohler2005a}
\begin{equation}
\label{eq:current}
\bar  I
= \frac{ge}{h}\int 
  d\epsilon \sum_{n}
  \big[T_{LR}^{(n)}(\epsilon)f_L(\epsilon)-T_{RL}^{(n)}(\epsilon)f_R(\epsilon)
  \big],
\end{equation}
where $e$ is the electron charge and $f(\epsilon)$ is the Fermi-Dirac
distribution. Spin and valley degeneracy of graphene is responsible for
the prefactor $g=4$, while $g=2$ accounts for the spin in a 2DEG. The
application of Eq.~\eqref{eq:current} to graphene (or any other
two-dimensional material) requires extending the summation to
transverse momenta $\hbar q$.  
For a driving field that breaks reflection symmetry, one generally
finds $T_{RL}(\epsilon) \neq T_{LR}(\epsilon)$.  Then, even when the
leads are in equilibrium such that $f_L(\epsilon)=
f_R(\epsilon)\equiv f(\epsilon)$, a net current may flow, and a pump current emerges,
\begin{equation}
\label{eq:current2}
\bar  I
= \frac{ge}{h}\int 
  d\epsilon f(\epsilon)\Delta T(\epsilon),
\end{equation}
where $\Delta T=T_{LR}-T_{RL}$.

\begin{figure}[tb]
   \centering
   \includegraphics[width=8.5cm]{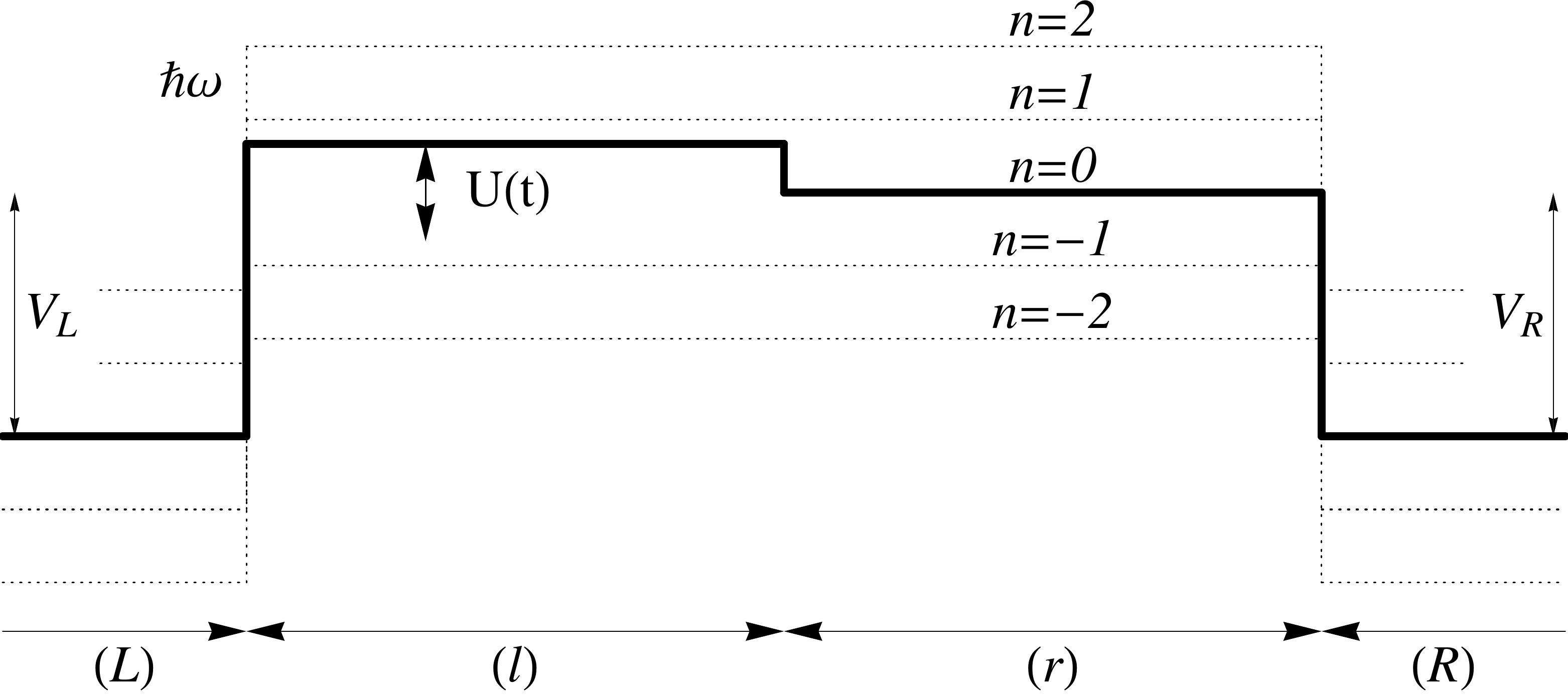}
   \caption{Quasi-one-dimensional scattering potential.
Profile of the pump, modeled as a barrier for which the left
half (region $l$) experiences an oscillatory gate voltage. Regions $L$
and $R$ correspond to highly doped leads. Dotted lines denote
sidebands to which the electron energy changes by absorption and
emission of photons from the driving field.}
   \label{Fig:potsketch}
\end{figure}

For the computation of the transmission probabilities we adopt the
Floquet scattering formalism of Ref.~\onlinecite{Wagner1995a} and
consider electrons in two dimensions under the influence of a
time-dependent potential described by the Hamiltonian
\begin{equation}
H(t) = H_0(x) + U(x)\cos(\omega t) .
\end{equation}
Here, $H_0$ comprises the kinetic energy and the static potential
$V(x)$, while $U(x)$ is the profile of the time-dependent potential
with frequency $\omega$.  Since the potential is $y$-independent, the
transverse momentum is conserved and the problem becomes effectively
one-dimensional.
We assume that both the static potential $V(x)$ and the driving
profile $U(x)$ are piecewise constant, and that $U(x)=0$ outside the
scattering region; see Fig.~\ref{Fig:potsketch}.  For graphene, $H_0 =
\hbar v_F\bm{k}\cdot\bm{\sigma} + V(x)$ with the wavevector $\bm{k} =
\pm k\bm e_x + q\bm e_y$.  The free solution with energy $E$ reads
\begin{equation}
\label{freeG}
\varphi_E^{\pm}
= \frac{e^{\pm ikx}}{\sqrt{2 |E| k/\hbar}}
  \begin{pmatrix} |E|/\hbar v_F \\ \pm k+iq\end{pmatrix} ,
\end{equation}
where $k$ is positive and fulfills the dispersion relation $E^2 =
\hbar^2v_F^2(k^2+q^2)$, while $\pm$ is the sign of the
corresponding current.  The normalization has been chosen such that
propagating waves have unit longitudinal current, $v_F
(\varphi_E^{\pm})^\dagger\sigma_x\varphi_E^{\pm} = \pm 1$.
This is convenient since then the coefficients of a superposition
become probability amplitudes.  For evanescent solutions with
imaginary longitudinal wavenumber $k=i\kappa$, the current vanishes.

According to the Floquet theorem, the Schr\"odinger equation with a
time-periodic Hamiltonian $H(t)=H(t+2\pi/\omega)$ possesses a complete set of
solutions with structure $\psi = e^{-i\epsilon t/\hbar} \phi(t)$, where
the Floquet state $\phi(t)=\phi(t+2\pi/\omega)$ obeys the time-periodicity of
the Hamiltonian and $\epsilon$ is the quasienergy.
Here we are looking for Floquet scattering states, i.e., solutions of
the Schr\"odinger equation that (i) are of Floquet structure and (ii)
have an incoming plane wave as boundary condition.  For clarity, we
derive here only the transmission from left to right---for the
opposite direction, it follows by simple re-labeling.

Condition (i) is equivalent to employing for the wavefunction
in any of the four regions $\ell = L,l,r,R$ the ansatz $\psi_\ell(x,t)
= e^{-i\epsilon t/\hbar}\phi_\ell(x,t)$.  The time-periodic parts
$\phi_\ell(x,t)$ still have to be determined, while the quasienergy
$\epsilon$ turns out to equal the energy of the incoming wave.
Inserting $\psi_\ell$ into the Schr\"odinger equation of region
$\ell$ yields for $\phi_\ell(x,t)$ a partial differential equation
which we solve by a separation ansatz.  The resulting solutions
\begin{eqnarray}\label{Floquetansatz}
\phi^\pm_{n,\ell}(x,t)
&=& e^{-in\omega t -i
  (U_\ell/\hbar\omega)
  \sin(\omega t)}
  \varphi^\pm_{\epsilon+n\hbar\omega-V_\ell}(x)\\
&=& \sum_{n'=-\infty}^\infty J_{n'-n}\left({U_\ell}/{\hbar \omega}\right)
  e^{-in'\omega t} \varphi^\pm_{\epsilon+n\hbar\omega-V_\ell}(x) ,\nonumber
\end{eqnarray}
comply with the requirement of time-periodicity provided that the
separation parameter $n$ is of integer value.
The separation parameter labels all possible solutions and determines
the time-averaged energy $\epsilon+n\hbar\omega$.  The Bessel
function of the first kind $J_n$ stems from the relation
$\exp[-iz\sin(\omega t)] = \sum_n J_n(z)\exp(-in\omega t)$.
Note that in the present case, $U_\ell$ is non-zero only in the
driving region $\ell=l$.
The ansatz \eqref{Floquetansatz} has also been used to study
photo-assisted tunneling in graphene.\cite{Trauzettel2007a,
*Trauzettel2007aE, AhsanZeb2008a, *AhsanZeb2008aE}

Condition (ii) means that in region $L$, the Floquet state consists
of an incoming plane wave and a reflected part, $\phi_L =
\varphi^+_{\epsilon-V_L} + \sum_n r_n\phi^-_{n,L} $, while in region
$R$, we have only an outgoing state, $\phi_R = \sum_n
t_n\phi^+_{n,R}$.
With the normalization chosen for the free solutions \eqref{freeG},
the coefficients of the latter superposition relate to the
left-to-right transmission probability under absorption or emission of $n$
quanta according to
\begin{equation}
T^{(n)}_{LR}(\epsilon) = |t_n(\epsilon)|^2 .
\end{equation}
In the scattering regions $l$ and $r$, the Floquet solution must be a
superposition of the states \eqref{Floquetansatz}, i.e.,
$\phi_\ell = \sum_n t_n^{(\ell)}\phi^+_{n,\ell} + \sum_n
r^{(\ell)}_n\phi^-_{n,\ell}$.
The coefficients $t_n^{(\ell)}$ and $r_n^{(\ell)}$ follow from the requirement
that the scattering states have to be continuous at any time.  These
matching conditions can be written as a set of linear equations, with
the incoming wave appearing as inhomogenity.  In matrix notation it reads
\begin{equation}
\label{wavematching}
\sum_{n'} M_{n,n'} \cdot
(r_{n'}, t^{(l)}_{n'}, r^{(l)}_{n'}, t^{(r)}_{n'}, r^{(r)}_{n'}, t_{n'} )
= \delta_{n,0}\tilde\varphi_{L,\mathrm{in}}^+ ,
\end{equation}
where $\tilde\varphi^+_{L,\mathrm{in}}\equiv\varphi^+_{\epsilon-V_L}$
denotes the incoming wave written as a 6-dimensional vector.  The
$6\times6$ matrices $M_{n,n'}$
are constructed from the wavefunctions \eqref{Floquetansatz} evaluated
at the interfaces.  They can be written efficiently as
\begin{equation}
M_{n,n'} = M(\epsilon-n'\hbar\omega) \sum_\ell J_{n'-n}\left({U_\ell}/{\hbar \omega}\right)P_\ell ,
\end{equation}
where $M(\epsilon)$ contains the wave matching condition at energy $\epsilon$
in the absence of driving.  The matrix $P_\ell$ is a projector to
region $\ell$, constructed such that $M P_\ell$ contains only
wavefunctions from region $\ell$.  In particular, $P_l =
\text{diag}(0,1,1,0,0,0)$ projects onto the driving region.
Equation~\eqref{wavematching} fully determines the
transmission and reflection amplitudes $t_n$ and $r_n$.
The tight-binding version\cite{Kohler2005a} of this method
is suitable for studying ac driving of smaller carbon-based
conductors, such as ribbons with a size of only a few lattice
constants\cite{Gu2009a} or nanotubes.\cite{FoaTorres2009a,FoaTorres2011a}
Via time-dependent density functional theory, it can be generalized to
the presence of interactions.\cite{Stefanucci2008a}

For electrons in a 2DEG with effective mass $m^*$, the static
Hamiltonian reads $H_0 = \hbar^2\bm{k}^2/2m^*+V(x)$.  Its free
solutions are scalar fields, but since also their derivatives must be
continuous, it is convenient to write them in spinor notation,
\begin{equation}
\label{freeN}
\varphi^{\pm}_E
= \frac{e^{\pm ikx}}{\sqrt{\hbar k/m^*}} \begin{pmatrix} 1 \\ \pm ik\end{pmatrix} ,
\end{equation}
with the dispersion relation $E=\hbar^2\bm{k}^2/2m^*$. The normalization is
such that the first vector component has unit current,
$\frac{1}{m^*}(\varphi_{E,1}^{\pm})^\dagger (-i\hbar\partial_x)
\varphi_{E,1}^{\pm} = \pm 1$.  With these ingredients, the Floquet
scattering formalism can be directly applied to the 2DEG case.



\subsection{Weak driving limit}\label{subsec:weakdriving}

The set of linear equations~\eqref{wavematching} can be solved
analytically in the limit of small driving amplitudes or large
frequencies, such that $p \equiv (U/2\hbar\omega)^2 \ll 1$.  We aim at finding
analytical expressions for the one-photon transmission probabilities
$T^{(\pm1)}$ in terms of the static transmission for Klein tunneling
at energy $\epsilon$ with respect to the top of a high barrier
\cite{Katsnelson2006a} of length $L$,
\begin{equation}
T(\epsilon,q)
= \frac{k_\epsilon^2}{k_\epsilon^2+q^2\sin^2(k_\epsilon L)},
\end{equation}
where $k_\epsilon=[(\epsilon/\hbar v_F)^2-q^2]^{1/2}$.
The semiclassical limit $L\to \infty$ of the above relation is
obtained by integration over fast oscillations, which yields
$T(\epsilon,q)\approx\sqrt{1-(\hbar v_Fq/\epsilon)^2}$.  In order to
relate these expressions to our formalism, we extract from
the solution of Eq.~\eqref{wavematching} in the undriven limit, $U=0$,
the transmission amplitude $t_0$ by multiplication with the vector
$\mathrm{p}_R^\dagger = (0,0,0,0,0,1)$ and obtain
\begin{equation}
\label{Tstatic}
T(\epsilon) = |\mathrm{p}_R^\dagger M^{-1}(\epsilon) \tilde\varphi_{L,\mathrm{in}}^+|^2.
\end{equation}

Next we simplify Eq.~\eqref{wavematching} using for the Bessel
functions the approximations $J_0(U/\hbar\omega)=1$,
$J_{\pm1}(U/\hbar\omega)=\pm U/2\hbar\omega$, while
$J_n(U/\hbar\omega) = \mathcal{O}^{|n|}\left(U/\hbar\omega\right)$ for
$U/\hbar\omega\ll1$.  Thus, to
first order in $U/\hbar\omega$, only the Floquet indices $n=0,\pm 1$ remain.
The resulting $18\times18$ matrix $(M_{n,n'})$ can be inverted via the
approximation $[A+(U/2\hbar\omega)B]^{-1} = A^{-1}- (U/2\hbar\omega)A^{-1}B
A^{-1} $, which provides $t_{\pm1}$ and, thus,
\begin{equation} \label{Tpm1} T^{(\pm1)}_{LR} = p
\left|\mathrm{p}_R^\dagger M^{-1}(\epsilon\mp\hbar\omega) M(\epsilon)
P_l M^{-1}(\epsilon) \tilde\varphi_{L,\mathrm{in}}^+\right|^2 .
\end{equation} The calculation of the elastic transmission
$T^{(0)}_{LR}$ up to leading order in $p$ is more tedious, but
fortunately not required, because our Hamiltonian obeys time-reversal
symmetry.  Therefore, $T^{(0)}_{LR}(\epsilon) =
T^{(0)}_{RL}(\epsilon)$ which implies that the elastic channel does
not contribute to the pump current.\cite{Kohler2005a}

Our numerical calculations will reveal that the relevant contributions
to the pump current stem from modes that, in the absence of driving,
are evanescent in the barrier region. For such modes, Eq.~\eqref{Tpm1}
can be connected to Eq.~\eqref{Tstatic} in a simple way, because the
modified source term in Eq.~\eqref{Tpm1} has an invariant
\emph{forward} amplitude, $M(\epsilon)P_lM^{-1}(\epsilon)
\tilde\varphi^+_{L,\mathrm{in}} =\tilde\varphi^+_{L,\mathrm{in}}
+\alpha^-\tilde\varphi^-_{L,\mathrm{in}}$.
Although $\alpha^-$ is not necessarily zero, it does not affect the
transmission, since it merely redefines the reflection amplitude
$r_0$. As a result, the one-photon transmission, Eq.~\eqref{Tpm1}, becomes,
besides a prefactor $p$, identical to the static transmission
\eqref{Tstatic} in the evanescent region,
\begin{equation}
\label{Tpm1T}
T^{(\pm1)}_{LR}(\epsilon) = p\, T(\epsilon\pm\hbar\omega,q).
\end{equation}
For the evanescent modes entering from the right, the same reasoning
lets us conclude that the waves decay exponentially and do not reach
the driven region $l$.  Consequently, $P_l M^{(-1)}(\epsilon)
\tilde\varphi^-_{R,\mathrm{in}} \approx 0$, and
$T^{(\pm1)}_{RL}(\epsilon) \ll p$ can be neglected.  Then the net
transmission appearing in Eq.~\eqref{eq:current2} becomes
\begin{equation}
\Delta T=p
  \left[T(\epsilon+\hbar\omega,q) +T(\epsilon-\hbar\omega,q)\right]
  +\mathcal{O}(e^{-2\kappa L}) .
\label{Tev}
\end{equation}
The magnitude of the correction reflects the fact that we have neglected
the exponentally small transmission of the evanescent modes which
decays with the imaginary wavenumber $i\kappa = i[q^2-(\epsilon/\hbar
v_F)^2]^{1/2}$.  Since $0\leq T \leq 1$, it follows that the maximal
net transmission for evanescent waves is $2p$.


\section{Non-adiabatic pump current}\label{current}

\begin{figure} 
   \centering
   \includegraphics[width=8.5cm]{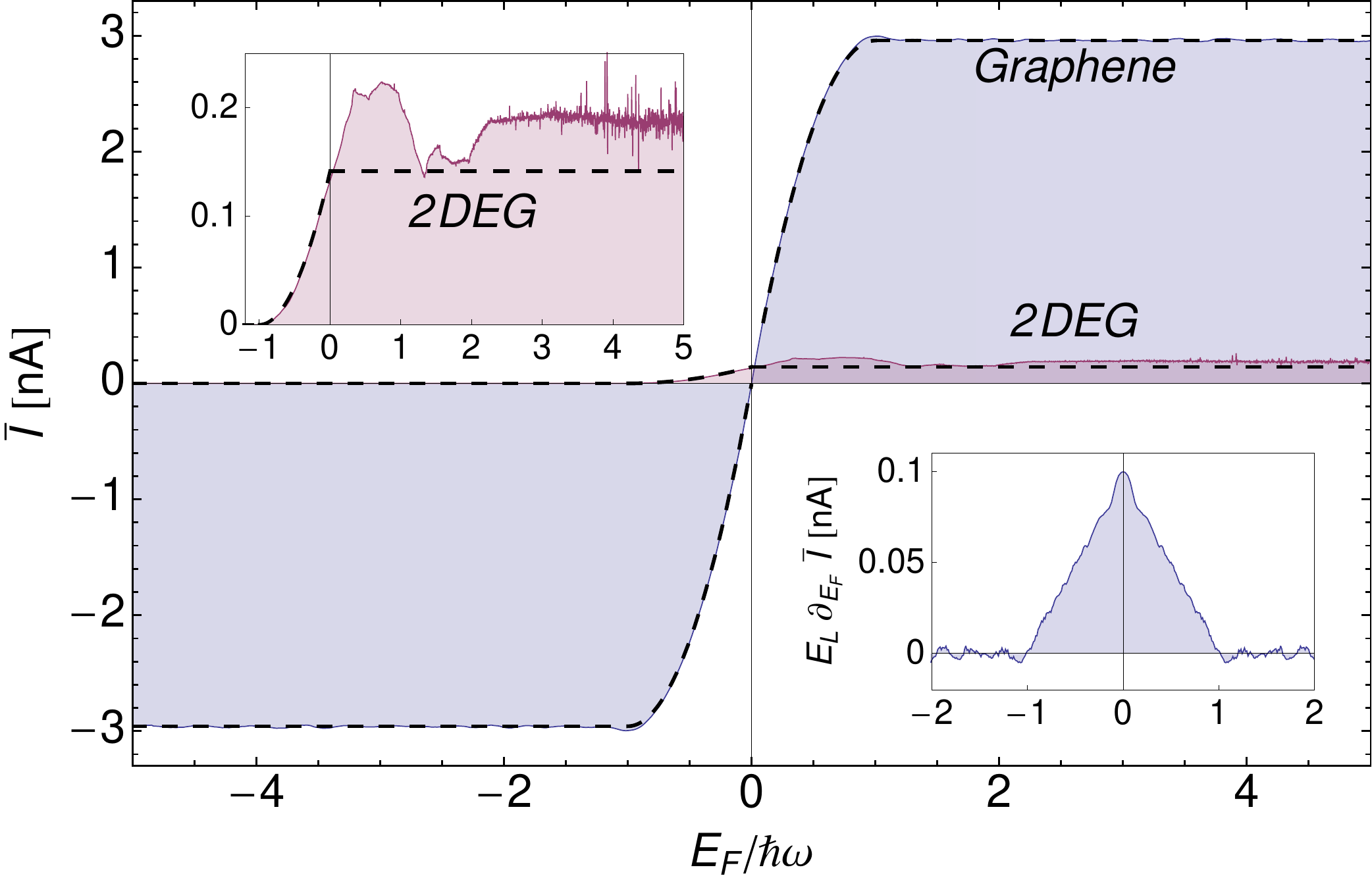}
   \caption{Pump current $\bar I$ for graphene and 2DEG. Dashed lines
mark the semiclassical result, which closely matches the numerical
ones. The upper inset is a blowup of the 2DEG case. The lower inset
depicts the differential current demonstrating that the main
contribution to the pump current arises from an energy range
$\hbar\omega$ around the Dirac point, populated by modes that are
evanescent in the barrier region.}
   \label{Fig:Sigma}
\end{figure}

In the relevant weak driving regime $U\ll\hbar \omega$, the absorption
or emission probabilities are of the order $p\equiv(U/2\hbar\omega)^2
\ll 1$ [see Eq.\ \eqref{Tpm1T}], so that $\Delta T\sim p$. Moreover,
for short and wide systems (width $W\gg L$), the total current takes
the form of an integral over modes $q$. Hence, the current \eqref{eq:current2} at a given
Fermi energy $E_F$ (measured from the Dirac point of the barrier) can be expressed as
\begin{equation}
\bar I = \frac{ge}{h}\left(\frac{U}{2\hbar\omega}\right)^2 W \int^{E_F}_{-\infty} d\epsilon
\int_{-\infty}^{\infty} dq ~ \frac{\Delta T}{p}.
\label{eq:sigma}
\end{equation}
We compute $\Delta T$ numerically by wave-matching in Floquet space
[see Eq.\ \eqref{wavematching}]. Using typical parameters
$L=5\mathrm{\mu m}$, $W/L=4$, $U=40\mu$eV,
$\hbar\omega=2$meV (around $500$GHz), we obtain the results shown in
Fig.~\ref{Fig:Sigma}.  As long as $p\ll 1$, the structure of the
transmission for graphene depends only on the product $L\omega$,
unlike for the 2DEG.  Graphene develops a far larger pump current
than a 2DEG, saturating to an $L$- and $\omega$-independent maximal
value $\bar I^\mathrm{max}\approx 3$nA for
$|E_F|\geq\hbar\omega$, while the smaller pump current in the 2DEG
case already saturates at $|E_F|\gtrsim 0$. The semiclassical
approximation described in Sec.~\ref{subsec:weakdriving}, valid for
$\hbar \omega\gg E_L$, yields
\begin{equation}
\label{IG}
\bar I_G \approx
  \frac{e}{\hbar}\frac{(U/2)^2}{E_W^G} \times
\begin{cases}
   \left(2-\frac{|E_F|}{\hbar\omega}\right)\frac{E_F}{\hbar\omega} ,
 & |E_F|<\hbar\omega \\
  \pm 1 ,
 & |E_F|>\hbar\omega
\end{cases} ,
\end{equation}
where the energy $E_W$ is the analogue of $E_L$ with $L$ replaced by the width $W$.
%
The corresponding calculation for a 2DEG provides the result
\begin{equation*}
\bar I_N \approx
  \frac{e}{\hbar}\frac{(U/2)^2}{2 k_F^{(\infty)}W E_W^N} \times
\begin{cases}
   0, & E_F< -\hbar\omega \\
   \left(1+\frac{E_F}{\hbar\omega}\right)^2 , & -\hbar\omega < E_F< 0 \\
   1 , & E_F>0
\end{cases}.
\end{equation*}
These approximations are plotted as dashed
curves in Fig.~\ref{Fig:Sigma}. Unlike for graphene, the pump current in the 2DEG depends on the leads' large Fermi momentum $k^{(\infty)}_F $ (around
$12\,\mathrm{nm}^{-1}$ for gold electrodes), which has been assumed
equal for both leads. The relative pump performance
$\nu\equiv {\bar I_G^\mathrm{max}}/{\bar I_N^\mathrm{max}}
=\hbar k^{(\infty)}_F/m^*v_F$, assuming gold
electrodes and GaAs/AlGaAs 2DEGs, is $\nu\approx 20$, i.e., the pump
current in the graphene device is much larger than in the 2DEG device.

\subsection{Direction-dependent excitation of evanescent modes}\label{direction}

In the following we show that the large and robust pump current of
the graphene device stems from a mechanism whereby the ac field
promotes evanescent modes from the left lead with probability $p$
into propagating modes, unlike evanescent modes from the right lead, which
couple poorly to the driving region.  In order to support this
picture, we consider the differential response $d\bar I/d E_F$,
shown in the lower inset of Fig.~\ref{Fig:Sigma}. It indicates that the main
contribution to the current stems from the \emph{bipolar regime}
$|\epsilon|<\hbar\omega$ in the case of graphene, or from the gap boundary
region $-\hbar\omega<\epsilon<0$ in the case of the 2DEG. In the absence of
driving, these energy ranges are populated by electronic modes that become
evanescent in the barrier region. Whether modes are evanescent or propagating
furthermore depends on their transverse momentum $q$.
Their response to driving is 
encoded in the function $\Delta T/p$, which represents the
differential response for a given mode with transverse momentum $q$
and total energy $\epsilon$ [see Eq.~\eqref{eq:sigma}].

\begin{figure}
   \centering
   \includegraphics[width=8.5cm]{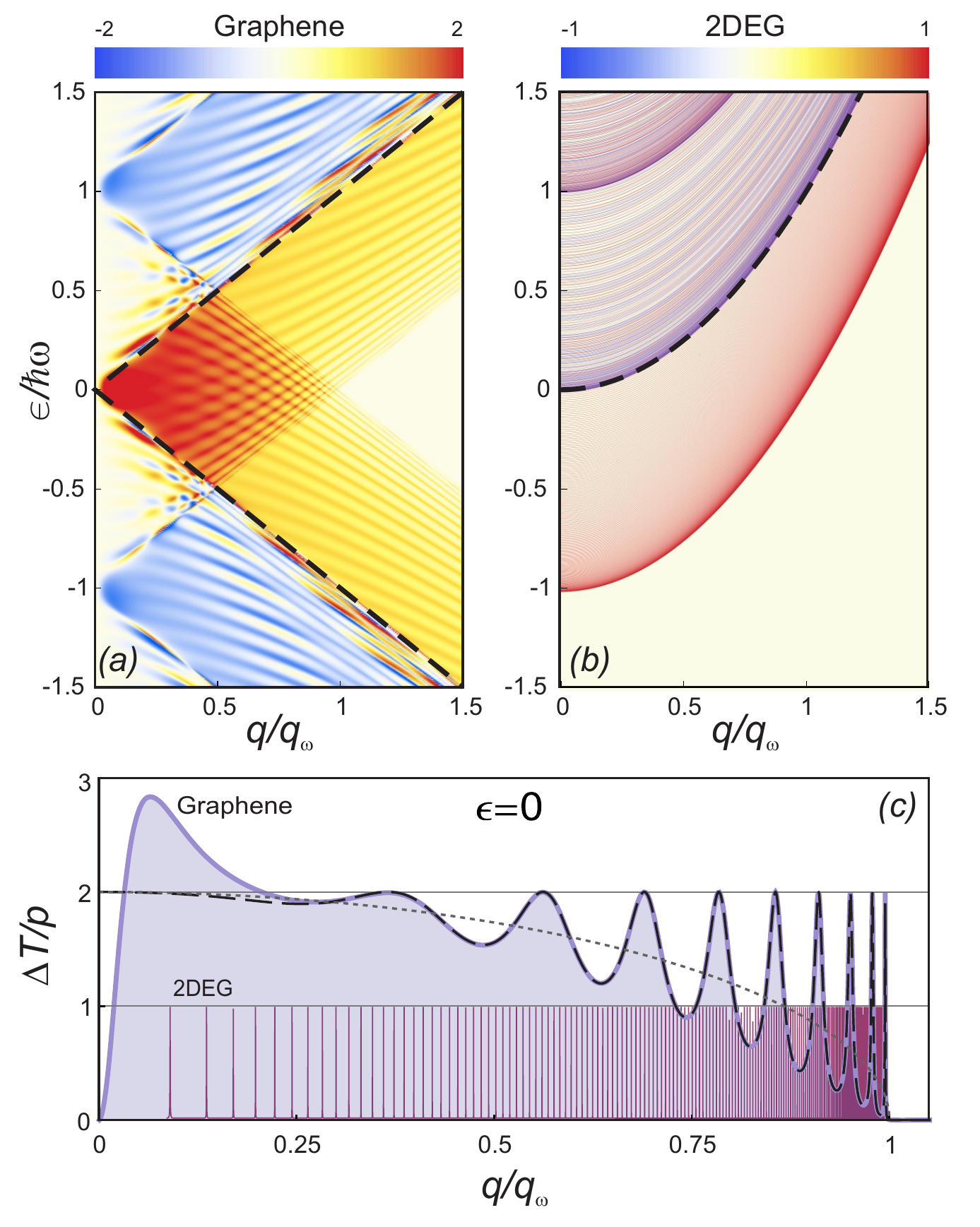}
   \caption{
Scaled net transmission $\Delta T/p$, where $p=(U/2\hbar\omega)^2$,
for (a) the graphene and (b) the 2DEG pump as a function of transverse
wave number $q$ and the initial energy $\epsilon$ for weak driving
$U\ll\hbar\omega$ at photon energy $\hbar\omega=2\mathrm{meV} \sim
500\mathrm{GHz}$. The dashed lines separate regions with propagating
and evanescent modes under the barrier.  (c) Cut at zero energy
revealing the behavior in the evanescent region. The dashed line marks
the static transmission $2T$ for graphene at energy $\hbar\omega$,
while the dotted curve is its semiclassical approximation. The
transverse wave number $q$ is scaled by $q_\omega$, which is defined
by $q_\omega L=\hbar\omega/E_L^G$ for graphene, and $(q_\omega L)^2 =
\hbar\omega/E_L^N$ for the 2DEG.}
   \label{Fig:DeltaT}
\end{figure}

Figure~\ref{Fig:DeltaT} depicts $\Delta T/p$ for graphene and the
2DEG. In the 2DEG case, at each energy only a discrete set of
\emph{resonant modes} contributes to the pump effect. These
resonances correspond, up to a shift of $\pm\hbar\omega$ in energy, to
quasibound levels in the static system,
that form because of the velocity mismatch at the interface with the metallic
leads, which results in strong confinement. The contact resistance increases
with increasing Fermi momentum $k_F^{(\infty)} L$, and the resonances become
increasingly narrow, resulting in a suppressed response of the 2DEG.  By
contrast, in graphene a broad range of modes contributes at all energies, and
the response is particularly strong in a diamond-shaped region within the
bipolar regime (red square in Fig.~\ref{Fig:DeltaT}), in agreement
with our earlier observation in Fig.~\ref{Fig:Sigma}. Graphene's
response does not exhibit sharp resonances due to the fact that, in
the static case, carrier chirality prohibits their confinement, so
that carriers in the graphene pump remain strongly coupled to the
leads, even when they are driven far out of equilibrium. Instead,
graphene's response in Fig.~\ref{Fig:DeltaT}(a) has
features associated with the cone $\epsilon=\hbar v_F q$ (dashed lines) and
its replicas shifted by multiples of $\hbar\omega$.  For the weak driving
considered here, only the two replicas at $\epsilon=\hbar v_F
q\pm\hbar\omega$ are visible.  The first cone divides the modes
entering the scattering region into two categories, propagating and
evanescent: any incoming carrier in mode $q$ will become evanescent
under the barrier if its energy fulfills $|\epsilon|<\hbar v_F q$, and
will remain propagating otherwise. The other two cones determine
whether carriers populate evanescent or propagating modes after
absorption or emission of a single photon. Analogously, in the 2DEG
the threshold between propagating and evanescent modes is found at
$\epsilon=\hbar^2 q^2/2m^*$, and under photon emission or absorption
shifts to $\epsilon=\hbar^2 q^2/2m^*\pm\hbar\omega$.

\begin{figure}[tb]
   \centering
   \includegraphics[width=8.5cm]{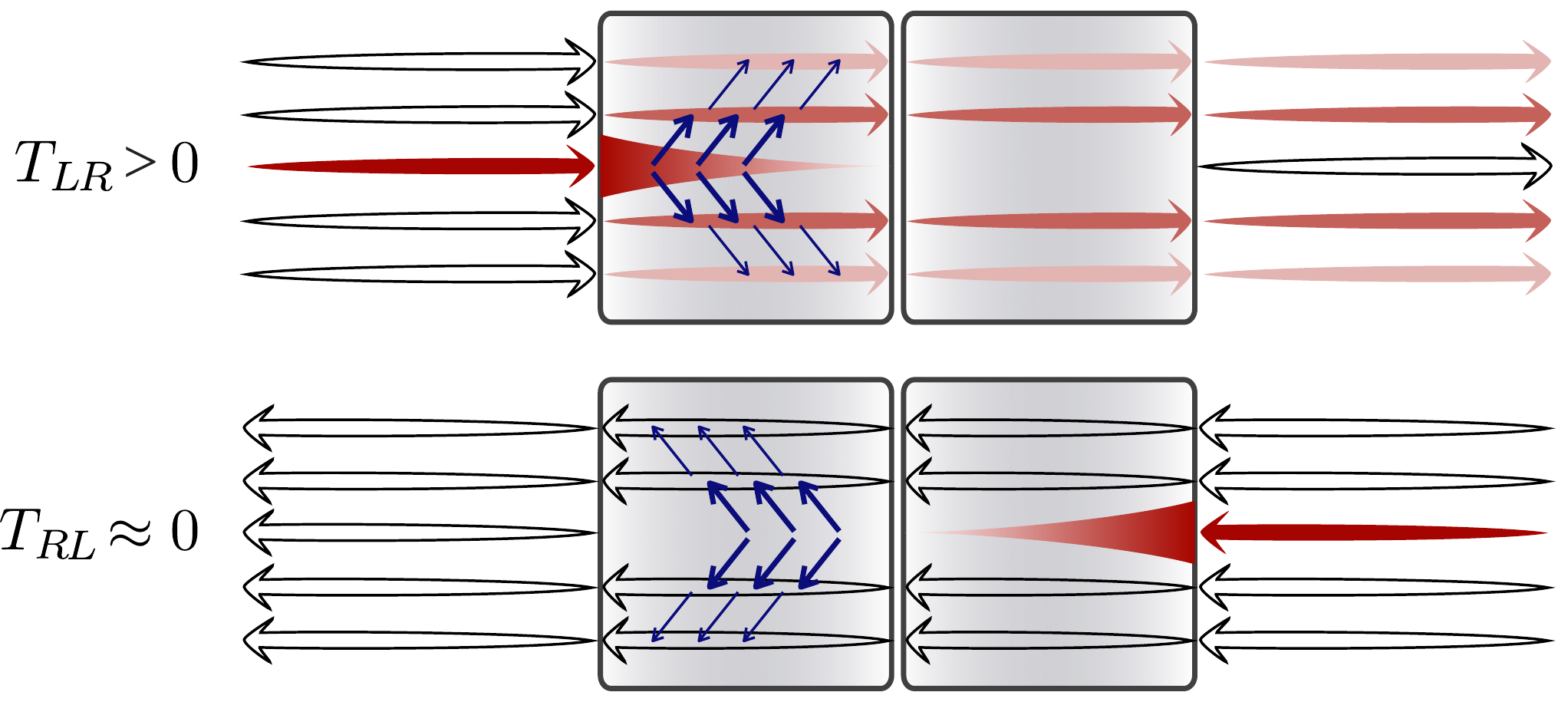}
   \caption{Evanescent mode pump mechanism.  Evanescent modes
penetrating the barrier from the left absorb or emit photons such
that they become propagating.  The corresponding modes from the right
lead decay before reaching the region with the ac gating.  Filled-in arrows
indicate strong population; non-filled arrows mark negligible scattering
channels.}
   \label{Fig:evmodes}
\end{figure}

In both systems, it can now be seen that the main mechanism of charge
transfer is established by a process whereby an evanescent mode
coming from the left lead may get transmitted to the right by
absorbing or emitting a photon, as long as in this process it jumps to
an existing propagating mode that may travel into the right lead. An
evanescent coming from the right, in contrast, first
encounters the static region not covered by the gate, and so gets
reflected with a high probability (see Fig.~\ref{Fig:evmodes}). This
spatial asymmetry rectifies the carrier flux such that a net
transport from left to right emerges.

In the 2DEG, when photon absorption is at resonance with a quasibound state,
this process via evanescent modes yields a maximal value $\Delta T/p\sim
1$, but because the resonances are narrow the integrated current remains
small. In graphene, on the other hand, all evanescent modes with $q
L<\hbar\omega/E^G_L$ exhibit a response $\Delta T/ p\lesssim 2$, close to
\emph{twice} the optimal value  of resonant evanescent modes in a 2DEG, but
without the need to satisfy any resonance condition. The factor two comes from
the two possible transitions to a propagating mode, by absorption but also by
emission of a photon, i.e., it is directly related to bipolarity. Resonant
conditions are not required because Klein tunneling in the propagating modes
and macroscopic tunneling in evanescent modes keep the contacts always
open; this is directly related to chirality. In combination, these two
graphene-specific effects maximize the differential pump response around
the Dirac point, and thus determine the robust characteristic features of the
total pump current in Fig.~\ref{Fig:Sigma}.


\section{Conclusions}\label{conclusions}

With this work, we have put forward a pump mechanism that makes use of the
evanescent modes under a potential barrier in graphene.  If an ac gate
voltage acts upon the, say, left half of the barrier, electrons penetrating
the barrier from the left lead are excited into propagating modes and, thus,
will be scattered to the right lead.  By contrast, electrons from the
opposite side will reach the driving region only with exponentially small
probability.  This breaking of spatio-temporal symmetries induces
surprisingly large non-adiabatic pump currents in the range of
several nA, significantly larger than what has been observed so far
with semiconductors.  The main reason for this efficiency is that a
whole range of evanescent modes is excited, while in the corresponding
setup with a 2DEG, only isolated resonances contribute.  Despite the
large pump currents, the main effect stems from single-photon
absorption, which allows one to obtain analytical results. Moreover,
owing to current conservation, ratchets which are built by several pumps in
series, exhibit identical operation characteristics (repetition of the
element is therefore not desirable for practical implementations).

A further important observation is that the resulting
pump current changes smoothly, symmetrically, and almost monotonically
with the distance of the Fermi energy to the Dirac point. This
behaviour is a direct consequence of bipolarity and, thus, particular
to graphene pumps.  As an application, it allows steering the dc
current into a direction of choice by simply shifting the barrier height
via a local dc gate voltage across the Dirac point.  This effect can
also be used to detect a static electric field by measuring the
current in response to a small oscillating probe electric field, or
detecting the amplitude of such an oscillating field when the static
field is fixed to a moderately large value, beyond which the
response flattens out.

\acknowledgments

We acknowledge support by the CSIC JAE-Doc program and the Spanish
Ministry of Science and Innovation through grants FIS2008-00124/FIS
(P.S.-J), FIS2009-08744 (E.P.) and  MAT2008-02626 (S.K.).


\bibliography{biblio,new}

\begin{thebibliography}{28}%
\makeatletter
\providecommand \@ifxundefined [1]{%
 \@ifx{#1\undefined}
}%
\providecommand \@ifnum [1]{%
 \ifnum #1\expandafter \@firstoftwo
 \else \expandafter \@secondoftwo
 \fi
}%
\providecommand \@ifx [1]{%
 \ifx #1\expandafter \@firstoftwo
 \else \expandafter \@secondoftwo
 \fi
}%
\providecommand \natexlab [1]{#1}%
\providecommand \enquote  [1]{``#1''}%
\providecommand \bibnamefont  [1]{#1}%
\providecommand \bibfnamefont [1]{#1}%
\providecommand \citenamefont [1]{#1}%
\providecommand \href@noop [0]{\@secondoftwo}%
\providecommand \href [0]{\begingroup \@sanitize@url \@href}%
\providecommand \@href[1]{\@@startlink{#1}\@@href}%
\providecommand \@@href[1]{\endgroup#1\@@endlink}%
\providecommand \@sanitize@url [0]{\catcode `\\12\catcode `\$12\catcode
  `\&12\catcode `\#12\catcode `\^12\catcode `\_12\catcode `\%12\relax}%
\providecommand \@@startlink[1]{}%
\providecommand \@@endlink[0]{}%
\providecommand \url  [0]{\begingroup\@sanitize@url \@url }%
\providecommand \@url [1]{\endgroup\@href {#1}{\urlprefix }}%
\providecommand \urlprefix  [0]{URL }%
\providecommand \Eprint [0]{\href }%
\providecommand \doibase [0]{http://dx.doi.org/}%
\providecommand \selectlanguage [0]{\@gobble}%
\providecommand \bibinfo  [0]{\@secondoftwo}%
\providecommand \bibfield  [0]{\@secondoftwo}%
\providecommand \translation [1]{[#1]}%
\providecommand \BibitemOpen [0]{}%
\providecommand \bibitemStop [0]{}%
\providecommand \bibitemNoStop [0]{.\EOS\space}%
\providecommand \EOS [0]{\spacefactor3000\relax}%
\providecommand \BibitemShut  [1]{\csname bibitem#1\endcsname}%
\let\auto@bib@innerbib\@empty
\bibitem [{\citenamefont {Reimann}(2002)}]{Reimann2002a}%
  \BibitemOpen
  \bibfield  {author} {\bibinfo {author} {\bibfnamefont {P.}~\bibnamefont
  {Reimann}},\ }\href@noop {} {\bibfield  {journal} {\bibinfo  {journal} {Phys.
  Rep.}\ }\textbf {\bibinfo {volume} {361}},\ \bibinfo {pages} {57} (\bibinfo
  {year} {2002})}\BibitemShut {NoStop}%
\bibitem [{\citenamefont {H\"anggi}\ and\ \citenamefont
  {Marchesoni}(2009)}]{Hanggi2009a}%
  \BibitemOpen
  \bibfield  {author} {\bibinfo {author} {\bibfnamefont {P.}~\bibnamefont
  {H\"anggi}}\ and\ \bibinfo {author} {\bibfnamefont {F.}~\bibnamefont
  {Marchesoni}},\ }\href@noop {} {\bibfield  {journal} {\bibinfo  {journal}
  {Rev. Mod. Phys.}\ }\textbf {\bibinfo {volume} {81}},\ \bibinfo {pages} {387}
  (\bibinfo {year} {2009})}\BibitemShut {NoStop}%
\bibitem [{\citenamefont {Brouwer}(1998)}]{Brouwer1998a}%
  \BibitemOpen
  \bibfield  {author} {\bibinfo {author} {\bibfnamefont {P.~W.}\ \bibnamefont
  {Brouwer}},\ }\href@noop {} {\bibfield  {journal} {\bibinfo  {journal} {Phys.
  Rev. B}\ }\textbf {\bibinfo {volume} {58}},\ \bibinfo {pages} {R10135}
  (\bibinfo {year} {1998})}\BibitemShut {NoStop}%
\bibitem [{\citenamefont {Switkes}\ \emph {et~al.}(1999)\citenamefont
  {Switkes}, \citenamefont {Marcus}, \citenamefont {Campman},\ and\
  \citenamefont {Gossard}}]{Switkes1999a}%
  \BibitemOpen
  \bibfield  {author} {\bibinfo {author} {\bibfnamefont {M.}~\bibnamefont
  {Switkes}}, \bibinfo {author} {\bibfnamefont {C.~M.}\ \bibnamefont {Marcus}},
  \bibinfo {author} {\bibfnamefont {K.}~\bibnamefont {Campman}}, \ and\
  \bibinfo {author} {\bibfnamefont {A.~C.}\ \bibnamefont {Gossard}},\
  }\href@noop {} {\bibfield  {journal} {\bibinfo  {journal} {Science}\ }\textbf
  {\bibinfo {volume} {283}},\ \bibinfo {pages} {1905} (\bibinfo {year}
  {1999})}\BibitemShut {NoStop}%
\bibitem [{\citenamefont {Vartiainen}\ \emph {et~al.}(2007)\citenamefont
  {Vartiainen}, \citenamefont {Mottonen}, \citenamefont {Pekola},\ and\
  \citenamefont {Kemppinen}}]{Vartiainen2007a}%
  \BibitemOpen
  \bibfield  {author} {\bibinfo {author} {\bibfnamefont {J.~J.}\ \bibnamefont
  {Vartiainen}}, \bibinfo {author} {\bibfnamefont {M.}~\bibnamefont
  {Mottonen}}, \bibinfo {author} {\bibfnamefont {J.~P.}\ \bibnamefont
  {Pekola}}, \ and\ \bibinfo {author} {\bibfnamefont {A.}~\bibnamefont
  {Kemppinen}},\ }\href@noop {} {\bibfield  {journal} {\bibinfo  {journal}
  {Appl. Phys. Lett.}\ }\textbf {\bibinfo {volume} {90}},\ \bibinfo {pages}
  {082102} (\bibinfo {year} {2007})}\BibitemShut {NoStop}%
\bibitem [{\citenamefont {Oosterkamp}\ \emph {et~al.}(1998)\citenamefont
  {Oosterkamp}, \citenamefont {Fujisawa}, \citenamefont {van~der Wiel},
  \citenamefont {Ishibashi}, \citenamefont {Hijman}, \citenamefont {Tarucha},\
  and\ \citenamefont {Kouwenhoven}}]{Oosterkamp1998a}%
  \BibitemOpen
  \bibfield  {author} {\bibinfo {author} {\bibfnamefont {T.~H.}\ \bibnamefont
  {Oosterkamp}}, \bibinfo {author} {\bibfnamefont {T.}~\bibnamefont
  {Fujisawa}}, \bibinfo {author} {\bibfnamefont {W.~G.}\ \bibnamefont {van~der
  Wiel}}, \bibinfo {author} {\bibfnamefont {K.}~\bibnamefont {Ishibashi}},
  \bibinfo {author} {\bibfnamefont {R.~V.}\ \bibnamefont {Hijman}}, \bibinfo
  {author} {\bibfnamefont {S.}~\bibnamefont {Tarucha}}, \ and\ \bibinfo
  {author} {\bibfnamefont {L.~P.}\ \bibnamefont {Kouwenhoven}},\ }\href@noop {}
  {\bibfield  {journal} {\bibinfo  {journal} {Nature (London)}\ }\textbf
  {\bibinfo {volume} {395}},\ \bibinfo {pages} {873} (\bibinfo {year}
  {1998})}\BibitemShut {NoStop}%
\bibitem [{\citenamefont {Blumenthal}\ \emph {et~al.}(2007)\citenamefont
  {Blumenthal}, \citenamefont {Kaestner}, \citenamefont {Li}, \citenamefont
  {Giblin}, \citenamefont {Janssen}, \citenamefont {Pepper}, \citenamefont
  {Anderson}, \citenamefont {Jones},\ and\ \citenamefont
  {Ritchie}}]{Blumenthal2007a}%
  \BibitemOpen
  \bibfield  {author} {\bibinfo {author} {\bibfnamefont {M.~D.}\ \bibnamefont
  {Blumenthal}}, \bibinfo {author} {\bibfnamefont {B.}~\bibnamefont
  {Kaestner}}, \bibinfo {author} {\bibfnamefont {L.}~\bibnamefont {Li}},
  \bibinfo {author} {\bibfnamefont {S.}~\bibnamefont {Giblin}}, \bibinfo
  {author} {\bibfnamefont {T.~J. B.~M.}\ \bibnamefont {Janssen}}, \bibinfo
  {author} {\bibfnamefont {M.}~\bibnamefont {Pepper}}, \bibinfo {author}
  {\bibfnamefont {D.}~\bibnamefont {Anderson}}, \bibinfo {author}
  {\bibfnamefont {G.}~\bibnamefont {Jones}}, \ and\ \bibinfo {author}
  {\bibfnamefont {D.~A.}\ \bibnamefont {Ritchie}},\ }\href@noop {} {\bibfield
  {journal} {\bibinfo  {journal} {Nature Phys.}\ }\textbf {\bibinfo {volume}
  {3}},\ \bibinfo {pages} {343} (\bibinfo {year} {2007})}\BibitemShut {NoStop}%
\bibitem [{\citenamefont {DiCarlo}\ \emph {et~al.}(2003)\citenamefont
  {DiCarlo}, \citenamefont {Marcus},\ and\ \citenamefont
  {Harris}}]{DiCarlo2003a}%
  \BibitemOpen
  \bibfield  {author} {\bibinfo {author} {\bibfnamefont {L.}~\bibnamefont
  {DiCarlo}}, \bibinfo {author} {\bibfnamefont {C.~M.}\ \bibnamefont {Marcus}},
  \ and\ \bibinfo {author} {\bibfnamefont {J.~S.}\ \bibnamefont {Harris},
  \bibfnamefont {Jr.}},\ }\href@noop {} {\bibfield  {journal} {\bibinfo
  {journal} {Phys. Rev. Lett.}\ }\textbf {\bibinfo {volume} {91}},\ \bibinfo
  {pages} {246804} (\bibinfo {year} {2003})}\BibitemShut {NoStop}%
\bibitem [{\citenamefont {Kaestner}\ \emph {et~al.}(2008)\citenamefont
  {Kaestner}, \citenamefont {Kashcheyevs}, \citenamefont {Hein}, \citenamefont
  {Pierz}, \citenamefont {Siegner},\ and\ \citenamefont
  {Schumacher}}]{Kaestner2008a}%
  \BibitemOpen
  \bibfield  {author} {\bibinfo {author} {\bibfnamefont {B.}~\bibnamefont
  {Kaestner}}, \bibinfo {author} {\bibfnamefont {V.}~\bibnamefont
  {Kashcheyevs}}, \bibinfo {author} {\bibfnamefont {G.}~\bibnamefont {Hein}},
  \bibinfo {author} {\bibfnamefont {K.}~\bibnamefont {Pierz}}, \bibinfo
  {author} {\bibfnamefont {U.}~\bibnamefont {Siegner}}, \ and\ \bibinfo
  {author} {\bibfnamefont {H.~W.}\ \bibnamefont {Schumacher}},\ }\href@noop {}
  {\bibfield  {journal} {\bibinfo  {journal} {Appl. Phys. Lett.}\ }\textbf
  {\bibinfo {volume} {92}},\ \bibinfo {pages} {192106} (\bibinfo {year}
  {2008})}\BibitemShut {NoStop}%
\bibitem [{\citenamefont {Fujiwara}\ \emph {et~al.}(2008)\citenamefont
  {Fujiwara}, \citenamefont {Nishiguchi},\ and\ \citenamefont
  {Ono}}]{Fujiwara2008a}%
  \BibitemOpen
  \bibfield  {author} {\bibinfo {author} {\bibfnamefont {A.}~\bibnamefont
  {Fujiwara}}, \bibinfo {author} {\bibfnamefont {K.}~\bibnamefont
  {Nishiguchi}}, \ and\ \bibinfo {author} {\bibfnamefont {Y.}~\bibnamefont
  {Ono}},\ }\href@noop {} {\bibfield  {journal} {\bibinfo  {journal} {Appl.
  Phys. Lett.}\ }\textbf {\bibinfo {volume} {92}},\ \bibinfo {pages} {042102}
  (\bibinfo {year} {2008})}\BibitemShut {NoStop}%
\bibitem [{\citenamefont {Kaestner}\ \emph {et~al.}(2009)\citenamefont
  {Kaestner}, \citenamefont {Leicht}, \citenamefont {Kashcheyevs},
  \citenamefont {Pierz}, \citenamefont {Siegner},\ and\ \citenamefont
  {Schumacher}}]{Kaestner2009a}%
  \BibitemOpen
  \bibfield  {author} {\bibinfo {author} {\bibfnamefont {B.}~\bibnamefont
  {Kaestner}}, \bibinfo {author} {\bibfnamefont {C.}~\bibnamefont {Leicht}},
  \bibinfo {author} {\bibfnamefont {V.}~\bibnamefont {Kashcheyevs}}, \bibinfo
  {author} {\bibfnamefont {K.}~\bibnamefont {Pierz}}, \bibinfo {author}
  {\bibfnamefont {U.}~\bibnamefont {Siegner}}, \ and\ \bibinfo {author}
  {\bibfnamefont {H.~W.}\ \bibnamefont {Schumacher}},\ }\href {\doibase
  10.1063/1.3063128} {\bibfield  {journal} {\bibinfo  {journal} {Appl. Phys.
  Lett.}\ }\textbf {\bibinfo {volume} {94}},\ \bibinfo {pages} {012106}
  (\bibinfo {year} {2009})}\BibitemShut {NoStop}%
\bibitem [{\citenamefont {Khrapai}\ \emph {et~al.}(2006)\citenamefont
  {Khrapai}, \citenamefont {Ludwig}, \citenamefont {Kotthaus}, \citenamefont
  {Tranitz},\ and\ \citenamefont {Wegscheider}}]{Khrapai2006a}%
  \BibitemOpen
  \bibfield  {author} {\bibinfo {author} {\bibfnamefont {V.~S.}\ \bibnamefont
  {Khrapai}}, \bibinfo {author} {\bibfnamefont {S.}~\bibnamefont {Ludwig}},
  \bibinfo {author} {\bibfnamefont {J.~P.}\ \bibnamefont {Kotthaus}}, \bibinfo
  {author} {\bibfnamefont {H.~P.}\ \bibnamefont {Tranitz}}, \ and\ \bibinfo
  {author} {\bibfnamefont {W.}~\bibnamefont {Wegscheider}},\ }\href@noop {}
  {\bibfield  {journal} {\bibinfo  {journal} {Phys. Rev. Lett.}\ }\textbf
  {\bibinfo {volume} {97}},\ \bibinfo {pages} {176803} (\bibinfo {year}
  {2006})}\BibitemShut {NoStop}%
\bibitem [{\citenamefont {Strass}\ \emph {et~al.}(2005)\citenamefont {Strass},
  \citenamefont {H\"anggi},\ and\ \citenamefont {Kohler}}]{Strass2005b}%
  \BibitemOpen
  \bibfield  {author} {\bibinfo {author} {\bibfnamefont {M.}~\bibnamefont
  {Strass}}, \bibinfo {author} {\bibfnamefont {P.}~\bibnamefont {H\"anggi}}, \
  and\ \bibinfo {author} {\bibfnamefont {S.}~\bibnamefont {Kohler}},\
  }\href@noop {} {\bibfield  {journal} {\bibinfo  {journal} {Phys. Rev. Lett.}\
  }\textbf {\bibinfo {volume} {95}},\ \bibinfo {pages} {130601} (\bibinfo
  {year} {2005})}\BibitemShut {NoStop}%
\bibitem [{\citenamefont {Klein}(1927)}]{Klein1929a}%
  \BibitemOpen
  \bibfield  {author} {\bibinfo {author} {\bibfnamefont {O.}~\bibnamefont
  {Klein}},\ }\href@noop {} {\bibfield  {journal} {\bibinfo  {journal} {Z.
  Phys.}\ }\textbf {\bibinfo {volume} {53}},\ \bibinfo {pages} {157} (\bibinfo
  {year} {1927})}\BibitemShut {NoStop}%
\bibitem [{\citenamefont {Katsnelson}\ \emph {et~al.}(2006)\citenamefont
  {Katsnelson}, \citenamefont {Novoselov},\ and\ \citenamefont
  {Geim}}]{Katsnelson2006a}%
  \BibitemOpen
  \bibfield  {author} {\bibinfo {author} {\bibfnamefont {M.~I.}\ \bibnamefont
  {Katsnelson}}, \bibinfo {author} {\bibfnamefont {K.~S.}\ \bibnamefont
  {Novoselov}}, \ and\ \bibinfo {author} {\bibfnamefont {A.~K.}\ \bibnamefont
  {Geim}},\ }\href@noop {} {\bibfield  {journal} {\bibinfo  {journal} {Nature
  Phys.}\ }\textbf {\bibinfo {volume} {2}},\ \bibinfo {pages} {620} (\bibinfo
  {year} {2006})}\BibitemShut {NoStop}%
\bibitem [{\citenamefont {Tworzydlo}\ \emph {et~al.}(2006)\citenamefont
  {Tworzydlo}, \citenamefont {Trauzettel}, \citenamefont {Titov}, \citenamefont
  {Rycerz},\ and\ \citenamefont {Beenakker}}]{Tworzydlo:PRL06}%
  \BibitemOpen
  \bibfield  {author} {\bibinfo {author} {\bibfnamefont {J.}~\bibnamefont
  {Tworzydlo}}, \bibinfo {author} {\bibfnamefont {B.}~\bibnamefont
  {Trauzettel}}, \bibinfo {author} {\bibfnamefont {M.}~\bibnamefont {Titov}},
  \bibinfo {author} {\bibfnamefont {A.}~\bibnamefont {Rycerz}}, \ and\ \bibinfo
  {author} {\bibfnamefont {C.~W.~J.}\ \bibnamefont {Beenakker}},\ }\href@noop
  {} {\bibfield  {journal} {\bibinfo  {journal} {Phys. Rev. Lett.}\ }\textbf
  {\bibinfo {volume} {96}},\ \bibinfo {pages} {246802} (\bibinfo {year}
  {2006})}\BibitemShut {NoStop}%
\bibitem [{\citenamefont {Prada}\ \emph {et~al.}(2009)\citenamefont {Prada},
  \citenamefont {San-Jose},\ and\ \citenamefont {Schomerus}}]{Prada:PRB09}%
  \BibitemOpen
  \bibfield  {author} {\bibinfo {author} {\bibfnamefont {E.}~\bibnamefont
  {Prada}}, \bibinfo {author} {\bibfnamefont {P.}~\bibnamefont {San-Jose}}, \
  and\ \bibinfo {author} {\bibfnamefont {H.}~\bibnamefont {Schomerus}},\
  }\href@noop {} {\bibfield  {journal} {\bibinfo  {journal} {Phys. Rev. B}\
  }\textbf {\bibinfo {volume} {80}},\ \bibinfo {pages} {245414} (\bibinfo
  {year} {2009})}\BibitemShut {NoStop}%
\bibitem [{\citenamefont {Kohler}\ \emph {et~al.}(2005)\citenamefont {Kohler},
  \citenamefont {Lehmann},\ and\ \citenamefont {H\"anggi}}]{Kohler2005a}%
  \BibitemOpen
  \bibfield  {author} {\bibinfo {author} {\bibfnamefont {S.}~\bibnamefont
  {Kohler}}, \bibinfo {author} {\bibfnamefont {J.}~\bibnamefont {Lehmann}}, \
  and\ \bibinfo {author} {\bibfnamefont {P.}~\bibnamefont {H\"anggi}},\
  }\href@noop {} {\bibfield  {journal} {\bibinfo  {journal} {Phys. Rep.}\
  }\textbf {\bibinfo {volume} {406}},\ \bibinfo {pages} {379} (\bibinfo {year}
  {2005})}\BibitemShut {NoStop}%
\bibitem [{\citenamefont {Wagner}\ and\ \citenamefont
  {Sols}(1999)}]{Wagner1999a}%
  \BibitemOpen
  \bibfield  {author} {\bibinfo {author} {\bibfnamefont {M.}~\bibnamefont
  {Wagner}}\ and\ \bibinfo {author} {\bibfnamefont {F.}~\bibnamefont {Sols}},\
  }\href@noop {} {\bibfield  {journal} {\bibinfo  {journal} {Phys. Rev. Lett.}\
  }\textbf {\bibinfo {volume} {83}},\ \bibinfo {pages} {4377} (\bibinfo {year}
  {1999})}\BibitemShut {NoStop}%
\bibitem [{\citenamefont {Wagner}(1995)}]{Wagner1995a}%
  \BibitemOpen
  \bibfield  {author} {\bibinfo {author} {\bibfnamefont {M.}~\bibnamefont
  {Wagner}},\ }\href@noop {} {\bibfield  {journal} {\bibinfo  {journal} {Phys.
  Rev. A}\ }\textbf {\bibinfo {volume} {51}},\ \bibinfo {pages} {798} (\bibinfo
  {year} {1995})}\BibitemShut {NoStop}%
\bibitem [{\citenamefont {Trauzettel}\ \emph {et~al.}(2007)\citenamefont
  {Trauzettel}, \citenamefont {Blanter},\ and\ \citenamefont
  {Morpurgo}}]{Trauzettel2007a}%
  \BibitemOpen
  \bibfield  {author} {\bibinfo {author} {\bibfnamefont {B.}~\bibnamefont
  {Trauzettel}}, \bibinfo {author} {\bibfnamefont {Y.~M.}\ \bibnamefont
  {Blanter}}, \ and\ \bibinfo {author} {\bibfnamefont {A.~F.}\ \bibnamefont
  {Morpurgo}},\ }\href@noop {} {\bibfield  {journal} {\bibinfo  {journal}
  {Phys. Rev. B}\ }\textbf {\bibinfo {volume} {75}},\ \bibinfo {pages} {035305}
  (\bibinfo {year} {2007})}\BibitemShut {NoStop}%
\bibitem [{\citenamefont {Trauzettel}\ \emph {et~al.}(2011)\citenamefont
  {Trauzettel}, \citenamefont {Blanter},\ and\ \citenamefont
  {Morpurgo}}]{Trauzettel2007aE}%
  \BibitemOpen
  \bibfield  {author} {\bibinfo {author} {\bibfnamefont {B.}~\bibnamefont
  {Trauzettel}}, \bibinfo {author} {\bibfnamefont {Y.~M.}\ \bibnamefont
  {Blanter}}, \ and\ \bibinfo {author} {\bibfnamefont {A.~F.}\ \bibnamefont
  {Morpurgo}},\ }\href@noop {} {\bibfield  {journal} {\bibinfo  {journal}
  {Phys. Rev. B}\ }\textbf {\bibinfo {volume} {83}},\ \bibinfo {pages}
  {159902(E)} (\bibinfo {year} {2011})}\BibitemShut {NoStop}%
\bibitem [{\citenamefont {Zeb}\ \emph {et~al.}(2008)\citenamefont {Zeb},
  \citenamefont {Sabeeh},\ and\ \citenamefont {Tahir}}]{AhsanZeb2008a}%
  \BibitemOpen
  \bibfield  {author} {\bibinfo {author} {\bibfnamefont {M.~A.}\ \bibnamefont
  {Zeb}}, \bibinfo {author} {\bibfnamefont {K.}~\bibnamefont {Sabeeh}}, \ and\
  \bibinfo {author} {\bibfnamefont {M.}~\bibnamefont {Tahir}},\ }\href@noop {}
  {\bibfield  {journal} {\bibinfo  {journal} {Phys. Rev. B}\ }\textbf {\bibinfo
  {volume} {78}},\ \bibinfo {pages} {165420} (\bibinfo {year}
  {2008})}\BibitemShut {NoStop}%
\bibitem [{\citenamefont {Zeb}\ \emph {et~al.}(2009)\citenamefont {Zeb},
  \citenamefont {Sabeeh},\ and\ \citenamefont {Tahir}}]{AhsanZeb2008aE}%
  \BibitemOpen
  \bibfield  {author} {\bibinfo {author} {\bibfnamefont {M.~A.}\ \bibnamefont
  {Zeb}}, \bibinfo {author} {\bibfnamefont {K.}~\bibnamefont {Sabeeh}}, \ and\
  \bibinfo {author} {\bibfnamefont {M.}~\bibnamefont {Tahir}},\ }\href@noop {}
  {\bibfield  {journal} {\bibinfo  {journal} {Phys. Rev. B}\ }\textbf {\bibinfo
  {volume} {79}},\ \bibinfo {pages} {089903(E)} (\bibinfo {year}
  {2009})}\BibitemShut {NoStop}%
\bibitem [{\citenamefont {Gu}\ \emph {et~al.}(2009)\citenamefont {Gu},
  \citenamefont {Yang}, \citenamefont {Wang},\ and\ \citenamefont
  {Chan}}]{Gu2009a}%
  \BibitemOpen
  \bibfield  {author} {\bibinfo {author} {\bibfnamefont {Y.}~\bibnamefont
  {Gu}}, \bibinfo {author} {\bibfnamefont {Y.~H.}\ \bibnamefont {Yang}},
  \bibinfo {author} {\bibfnamefont {J.}~\bibnamefont {Wang}}, \ and\ \bibinfo
  {author} {\bibfnamefont {K.~S.}\ \bibnamefont {Chan}},\ }\href {\doibase
  10.1088/0953-8984/21/40/405301} {\bibfield  {journal} {\bibinfo  {journal}
  {J. Phys.: Condens. Matter}\ }\textbf {\bibinfo {volume} {21}},\ \bibinfo
  {pages} {405301} (\bibinfo {year} {2009})}\BibitemShut {NoStop}%
\bibitem [{\citenamefont {Foa~Torres}\ and\ \citenamefont
  {Cuniberti}(2009)}]{FoaTorres2009a}%
  \BibitemOpen
  \bibfield  {author} {\bibinfo {author} {\bibfnamefont {L.~E.~F.}\
  \bibnamefont {Foa~Torres}}\ and\ \bibinfo {author} {\bibfnamefont
  {G.}~\bibnamefont {Cuniberti}},\ }\href {\doibase 10.1016/j.crhy.2009.05.003}
  {\bibfield  {journal} {\bibinfo  {journal} {C. R. Physique}\ }\textbf
  {\bibinfo {volume} {10}},\ \bibinfo {pages} {297} (\bibinfo {year}
  {2009})}\BibitemShut {NoStop}%
\bibitem [{\citenamefont {Foa~Torres}\ \emph {et~al.}(2011)\citenamefont
  {Foa~Torres}, \citenamefont {Calvo}, \citenamefont {Rocha},\ and\
  \citenamefont {Cuniberti}}]{FoaTorres2011a}%
  \BibitemOpen
  \bibfield  {author} {\bibinfo {author} {\bibfnamefont {L.~E.~F.}\
  \bibnamefont {Foa~Torres}}, \bibinfo {author} {\bibfnamefont {H.~L.}\
  \bibnamefont {Calvo}}, \bibinfo {author} {\bibfnamefont {C.~G.}\ \bibnamefont
  {Rocha}}, \ and\ \bibinfo {author} {\bibfnamefont {G.}~\bibnamefont
  {Cuniberti}},\ }\href {\doibase 10.1063/1.3630025} {\bibfield  {journal}
  {\bibinfo  {journal} {Appl. Phys. Lett.}\ }\textbf {\bibinfo {volume} {99}},\
  \bibinfo {pages} {092102} (\bibinfo {year} {2011})}\BibitemShut {NoStop}%
\bibitem [{\citenamefont {Stefanucci}\ \emph {et~al.}(2008)\citenamefont
  {Stefanucci}, \citenamefont {Kurth}, \citenamefont {Rubio},\ and\
  \citenamefont {Gross}}]{Stefanucci2008a}%
  \BibitemOpen
  \bibfield  {author} {\bibinfo {author} {\bibfnamefont {G.}~\bibnamefont
  {Stefanucci}}, \bibinfo {author} {\bibfnamefont {S.}~\bibnamefont {Kurth}},
  \bibinfo {author} {\bibfnamefont {A.}~\bibnamefont {Rubio}}, \ and\ \bibinfo
  {author} {\bibfnamefont {E.~K.~U.}\ \bibnamefont {Gross}},\ }\href@noop {}
  {\bibfield  {journal} {\bibinfo  {journal} {Phys. Rev. B}\ }\textbf {\bibinfo
  {volume} {77}},\ \bibinfo {pages} {075339} (\bibinfo {year}
  {2008})}\BibitemShut {NoStop}%
\end{thebibliography}%

\end{document}